\documentclass[a4paper,11pt]{article}

\pdfoutput=1
\usepackage{jcappub}

\usepackage{graphicx}
\usepackage{longtable}
\usepackage{float}
\usepackage{dcolumn}
\usepackage{bm}
\usepackage{appendix}
\usepackage{multirow}
\usepackage{color}
\usepackage[utf8]{inputenc}
\usepackage{footmisc}
\usepackage{txfonts}
\usepackage{xcolor}
\usepackage{graphicx}
\usepackage{bm}
\usepackage{ulem}
\usepackage{multirow, array, float, tabularx}
\usepackage{amsmath}
\usepackage[colorlinks=true,linkcolor=blue,citecolor=blue]{hyperref}

\newcommand{\be}{\begin{equation}}
\newcommand{\ee}{\end{equation}}
\newcommand{\bea}{\begin{eqnarray}}
\newcommand{\eea}{\end{eqnarray}}
\newcommand{\bes}{\begin{subequations}}
	\newcommand{\ees}{\end{subequations}}

\newcommand{\bc}{\begin{center}}
	\newcommand{\ec}{\end{center}}

\usepackage{wasysym}

\begin{document}

\title{Revisiting Witten-O'Raifeartaigh Inflation for a Non-minimally Coupled Scalar Field}
\author[a]{F. B. M. dos Santos}\emailAdd{felipe.santos.091@ufrn.edu.br}

\author[a,b]{R. Silva}
\emailAdd{raimundosilva@fisica.ufrn.br}

\affiliation[a]{Universidade Federal do Rio Grande do Norte,
	Departamento de F\'{\i}sica, Natal - RN, 59072-970, Brasil}

\affiliation[b]{Departamento de F\'{\i}sica, Universidade do Estado do Rio Grande do Norte, Mossor\'o, 59610-210, Brasil}

\abstract{
In this work, we revisit the Witten-O'Raifeartaigh model of inflation, in which the potential takes a $\operatorname{log}^2(\phi/M)$ form, when the scalar field is non-minimally coupled to gravity. We investigate the impact of the coupling in the prediction of the inflationary parameters, thereby affecting the viability of the model. We find that a small coupling of order $\xi\sim10^{-3}$ is preferred by data at the $n_s-r$ plane level, and that the presence of a non-zero $\xi$ allows for a large interval of the mass scale $M$, in which it is possible to achieve a low tensor-to-scalar ratio. We also establish constraints imposed by a subsequent reheating era, in which its duration and temperature can be related to CMB observables, which in return, restricts the possible values for the $n_s$ and $r$ parameters.}

\maketitle

\section{Introduction}\label{sec1}

Since its conception, the idea of cosmic inflation \cite{Starobinsky:1980te,Guth1981,Linde:1981mu,Linde:1983gd} as a solution for problems of the Big Bang picture has been regarded to be perhaps the most accurate description of the very early universe. In this picture, the universe would have undergone a rapid expansion caused by an agent, usually taken as a scalar field called \textit{inflaton}, that dominated the energy density of the universe. Such a period is well described by the slow-roll regime, in which the potential function $V(\phi)$ is flat enough for accelerated expansion to take place. The discovery of the Cosmic Microwave Background (CMB), and dedicated experiments that followed \cite{Mather:1993ij,WMAP:2003syu,WMAP:2006bqn}, corroborate the idea of this rapid early expansion of the universe, preparing it to a state that makes possible the evolution to what it is today. This state is characterized by the \textit{reheating} era, in which the energy stored in the inflaton field is converted to new particles, including the ones in the standard model of particle physics \cite{Abbott:1982hn,Albrecht:1982mp,Greene:1997fu,Kofman:1994rk,Kofman:1997yn}. The data extracted from CMB has been helping us to verify inflation, where currently, we have access to extensive data provided by the Planck collaboration \cite{Planck:2015sxf,Aghanim:2018eyx,Planck:2018jri}, being essential to select possible scenarios.

On the theoretical side, much has been done. Over the last decades, many models, coming from a very diverse variety of more fundamental theories, have appeared \cite{Martin:2013tda}. Many of them are still viable options as the CMB constraints become more restrictive. For this reason, it has become more critical than ever to find ways to determine how many of these models are good candidates for a realistic description of the inflationary era \cite{Martin:2013nzq}. However, the more popular scenarios studied involve a scalar field minimally coupled to gravity. Some frameworks, on the other hand, consider a field non-minimally coupled to gravity, being a topic of discussion \cite{Lucchin:1985ip, Futamase:1987ua, Komatsu1999, Komatsu1998, Fakir1990, Faraoni:1998qx, Hertzberg:2010dc, Bezrukov2008, Linde:2011nh, Kaiser:2015usz}. From the observational standpoint, the presence of a non-minimal coupling might lead to better predictions for a given model, essentially reviving models that were once excluded by the CMB data \cite{Tenkanen:2017jih}. In recent years, analyses seeking to confront non-minimally coupled models with CMB data have been done (See Refs. \cite{Okada:2014lxa,Tenkanen:2017jih, Campista:2017ovq,Ferreira:2018nav, Bostan:2018evz,Reyimuaji:2020goi, Rodrigues:2021txa,Rodrigues:2020dod,Santos:2021lqh}), all of which presenting interesting results, in the sense that cosmological parameters can be into Planck constraints, and with the extra parameters of a given model generally being well constrained by these data.

In this work, we investigate a specific model, with potential $V(\phi)=V_0\log^2(\phi/M)$ \cite{Albrecht:1983ib}, usually called the Witten-O'Raifeartaigh (WR) model \cite{Martin:2013tda}. This function arises from a generalized form of the O'Raifeartaigh superpotential \cite{ORaifeartaigh:1975nky}, with the role of trying to solve the hierarchy problem through supersymmetry (SUSY). In the original work, the mass scale $M$ was constrained to be the order of the Planck mass $M_p$ \cite{Albrecht:1983ib}, so the model in principle has only the parameter $V_0$ to be determined. In the analysis of the model described in \cite{Martin:2013tda} however, the model was treated in a phenomenological manner by allowing $M$ to change. In this regard, it greatly impacts the predictions of the inflationary parameters, where one can achieve a wide range of the tensor-to-scalar ratio value $r$. More recently, the same form of potential was derived from another supersymmetric construction \cite{Artymowski:2019jlh}, where there is no theoretical constraint on $M$. From an observational perspective, however, even by varying $M$, the predictions for the spectral index $n_s$ and the mentioned $r$ are only in the $2\sigma$ region of the most recent Planck confidence contours, for $N_\star=50-60$. Therefore, if one wishes to consider this scenario further to have better predictions, modifications at the level of the gravitational action might be needed. This way, we consider this model in the context of a field that is non-minimally coupled to gravity, where an extra parameter $\xi$ that controls the strength of the coupling appears. We will check if the presence of this coupling improves the predictions of the model initially by checking its compatibility with the $n_s$, $r$ and the running of the spectral index $n_{run}$ data from Planck. We then perform an analysis that relates quantities such as the reheating temperature $T_{re}$, its duration $N_{re}$, when the oscillations of the scalar field around the minimum can be parameterized by an equation of state parameter $\bar w_{re}$ \cite{Munoz:2014eqa,Cook:2015vqa,Cai:2015soa,Dai:2014jja,Ueno:2016dim,Eshaghi:2016kne,Kabir:2016kdh,DiMarco:2017zek,Drewes:2017fmn,Lopez:2021agu,Cheong:2021kyc}.

This work is organized in the following way: In Section \ref{sec2} we briefly describe the slow-roll picture for a non-minimally coupled field. In Section \ref{sec3}, we review the main theoretical motivations of the model and perform the slow-roll analysis in the presence of the non-minimal coupling. Section \ref{sec4} describes the method by which reheating and inflationary parameters are related, where we apply it to the WR model. Finally, in Section \ref{sec6}, we present our considerations.

\section{Non-Minimal Slow-roll inflation}\label{sec2}

To describe inflation when the field is non-minimally coupled to gravity, we start from the following action
\begin{equation}
S=\int d^4x\sqrt{-g}\left[\frac{\Omega^2(\phi)}{16\pi G}R - \frac{1}{2}g^{\mu\nu}\partial_\mu\phi\partial_\nu\phi - V(\phi)\right],
\label{1}
\end{equation}
known as the Jordan frame action, with $\Omega^2(\phi)=1+8\pi G\xi\phi^2$, and $\frac{1}{\sqrt{8\pi G}}=M_p$ being the reduced Planck mass. Also, $g^{\mu\nu}$ is considered to be the Friedman-Robertson-Lemaître-Walker (FLRW) metric. When $\xi=0$, this action reduces to that of a minimally-coupled scalar field; the parameter $\xi$ determines how strongly the inflaton is coupled to the gravity sector. In some models, e.g., Higgs inflation \cite{Bezrukov2008}, $\xi$ is constrained to large values, while in others, such as chaotic inflation, a small $\xi$, of order $\xi = \mathcal{O}(10^{-3})$ is enough to bring the spectral index and tensor-to-scalar ratio $n_s$ and $r$ back into the constraints imposed by the Planck measurements \cite{Tenkanen:2017jih}. Although we can work in the Jordan frame, it is usually simpler to work in the Einstein frame since the resulting equations of motion for the inflaton can be cast in the same form as for a minimally coupled scalar field, with the dependence on $\xi$ is explicit in the potential.

The Einstein frame description is achieved by a conformal transformation of the type $\tilde{g}_{\mu\nu}=\Omega^2(\phi)g_{\mu\nu}$, in which the field is redefined to a new one called $\chi(\phi)$. As a result, the action (\ref{1}) becomes
\begin{equation}
S = \int d^4x\sqrt{-g}\left[\frac{R}{16\pi G}-\frac{1}{2}\tilde{g}^{\mu\nu}\partial_\mu\chi\partial_\nu\chi-\hat{V}(\chi)\right],
\label{2}
\end{equation}
where 
\begin{equation}
\hat{V}(\chi)\equiv\frac{V(\phi)}{\left(1+\xi\frac{\phi^2}{M_p^2}\right)^2},
\label{3}
\end{equation}
is the Einstein frame potential, and the fields $\phi$ and $\chi$ are related by
\begin{equation}
\frac{d\chi}{d\phi}=\frac{\sqrt{1+\xi\frac{\phi^2}{M_p^2}(1+6\xi)}}{\left(1+\xi\frac{\phi^2}{M_p^2}\right)}.
\label{4}
\end{equation}
It is possible to take a limit on $\xi\phi^2/M_p^2$ and integrate (\ref{4}) analytically for $\phi$; this is useful when one wants to investigate the small or large coupling limits for a model. We choose, however, to leave $\xi$ unrestricted, so we use the potential $\hat{V}(\chi)$ as a function of $\phi$ instead, as in Eq. (\ref{3}). The slow-roll parameters are modified and are expressed as

\begin{gather}
\epsilon=\frac{M_p^2}{2}\left(\frac{V_{,\phi}}{V\chi_{,\phi}}\right),\quad \eta = M_p^2\left(\frac{V_{,\phi\phi}}{V\chi_{,\phi}^2} - \frac{V_{,\phi}\chi_{,\phi\phi}}{V\chi_{,\phi}^3}\right),\nonumber\\
\zeta^2 = M_p^4\frac{V_\phi}{V^2\chi_\phi^2}\left( \frac{V_{\phi\phi\phi}}{\chi_\phi^2} -\frac{3V_{\phi\phi}\chi_{\phi\phi}}{\chi_\phi^3} + \frac{3V_{\phi}\chi_{\phi\phi}^2}{\chi_\phi^4}-\frac{V_\phi\chi_{\phi\phi\phi}}{\chi_\phi^3} \right),
\label{5}
\end{gather} 

where  $_{,\phi}$ and $_{,\chi}$ mean a derivative with respect to $\phi$ and $\chi$, respectively, so that we still can impose $\epsilon,\vert\eta\vert\ll1$ when inflation happens. We can approximate the end of inflation when one of these conditions is violated, such as $\epsilon(\phi_{end})=1$, from which we can find the value of the field at the end of inflation $\phi_{end}$. We find the field $\phi_\star$ at horizon crossing from
\begin{equation}
N_\star=\frac{1}{M_p^2}\int_{\phi_{end}}^{\phi_{\star}}d\phi\frac{\hat V}{\hat V_\phi}\chi_{,\phi}^2.
\label{6}
\end{equation}
with $N_\star$ being the number of \textit{e-folds} at the same moment, usually taken as between $50-60$, for inflation to solve the flatness and horizon problems. One can fix the amplitude of a given scalar potential by using the amplitude of the power spectrum of scalar perturbations, computed when the CMB mode crosses the horizon at the reference pivot scale, that we take as $k_{\star}= 0.05$Mpc$^{-1}$
\begin{equation}
P_{R}=\frac{V(\phi_{\star})}{24\pi^2M_p^4\epsilon_\star}\Bigg|_{k=k_\star},
\label{7}
\end{equation}
in which the value of $P_{R}$ at the pivot scale is given by the \textit{Planck} collaboration as $\log(10^{10}P_{R})=3.044\pm0.014$ \cite{Planck:2018jri}. The tensor-to-scalar ratio $r$, the spectral index $n_s$ and its running are expressed in the usual manner as in the minimally coupled theory

\begin{equation}
r=16\epsilon, n_{s}=1-6\epsilon + 2\eta, \quad n_{run} = 16\epsilon\eta - 24\epsilon^2 - 2\zeta^2
\label{8}
\end{equation}
where $r$ has today an upper limit of $r<0.056$, while the spectral index is constrained as $n_s=0.9649 \pm 0.0042 $, at $68\%$ confidence level, for Planck TT,TE,EE+lowE+lensing \cite{Planck:2018jri} . To have a viable model, we must ensure that it predicts a low enough primordial gravitational wave spectrum and produces an almost scale-invariant scalar power spectrum.

\section{Non-minimal WR model}\label{sec3}

\begin{figure*}
	\centering
	\includegraphics[width=8cm]{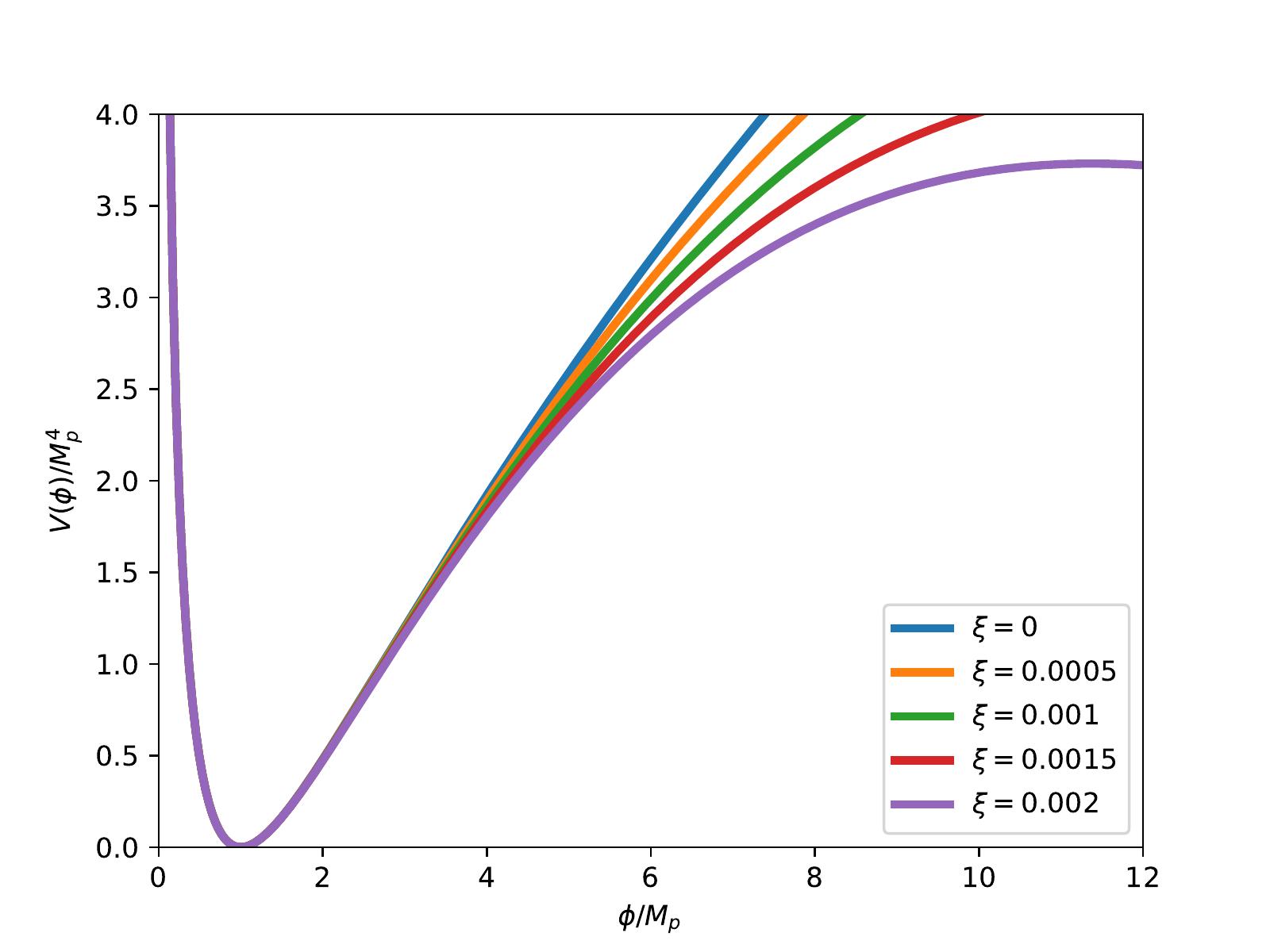}
	\caption{The Einstein frame potential (\ref{19}), for different values of $\xi$. We have set $M=M_p$.}
	\label{fig1}
\end{figure*}

This section reviews some theoretical motivations for the WR model of inflation. We also perform the slow-roll analysis of the non-minimally coupled version of this scenario through its dependence on the parameter $\xi$.

\subsection{Theoretical motivations}

For an inflationary model to be viable, one thing that is usually necessary is a potential with a region flat enough for the slow-roll regime to take place. A manner of obtaining such potentials is using the framework of SUSY and its local extension, or Supergravity (SUGRA) \cite{Ellis:1982ws,Ellis:1982dg,Albrecht:1983ib,Kallosh:2010ug,Yamaguchi:2011kg}. This is because, for a given number of fields, it is possible to find functions that possess flat directions that can be taken as the inflationary trajectory. One can derive many classes of potentials as part of a realistic inflationary scenario, given the needed ingredients, such as the Kähler function $K(T\bar T)$ and a superpotential $W(T)$, for fields $T,\bar T$, composing the called F and D-term potentials. However, a problem with the general construction of SUGRA inflationary models is an exponential dependence on the canonical $K=T\bar T$ in the F-term potential, causing it to be too steep for slow-roll inflation. Non-canonical forms of $K$ are used to solve this problem, such as $\alpha$-attractor models \cite{Kallosh:2013yoa,Kallosh:2013hoa,Kallosh:2013daa}, in which the appearance of non-canonical kinetic term results in a potential that has always a flat region in the large field limit \cite{Linde:2016uec}.

Here, we briefly review the original motivation for the WR model of inflation, which dates back to 
the first supersymmetric models in which SUSY is broken \cite{ORaifeartaigh:1975nky}, and it was also used as a way to address the hierarchy problem \cite{Witten:1981kv};
one can refer to \cite{Martin:2013tda} for a review. A popular form of the superpotential $W$ is the O'Raifeartaigh one 
\begin{equation}
W = \lambda X\left(A^2-m^2\right) + gYA,
\label{9}
\end{equation}
with $A, X$ and $Y$ being chiral superfields, $\lambda$ is a dimensionless constant, while $g,m$ have dimensions of mass. For the potential generated by (\ref{9}), supersymmetry is broken so that the potential receives a correction of a logarithm dependence on $X$. The presence of this term prevents $X$ from achieving larger values, thus making difficult for the model to solve the hierarchy problem \cite{Witten:1981kv}. On the other hand, from the point of view of non-Abelian theories, the one-loop corrections produced to the potential are necessarily negative, meaning that the potential is no longer stable and the field could go to very large values. In \cite{Witten:1981nf,Dimopoulos:1982gm}, a non-Abelian generalization of (\ref{9}) was investigated, in which the one-loop correction have the form $\log|X|^2$. This idea appeared in an inflationary model in \cite{Albrecht:1983ib}. As one-loop corrections to the generalized potential have the form $V\propto 1+\tilde b\log\frac{\phi}{\mu}$, where $\tilde b$ is in principle a constant, that if negative, leads the field towards $\phi\gg\mu$, and higher order corrections might produce a logarithm dependence of $\tilde b$ on $\phi$, and as a result, the potential acquires a minimum at $\phi=M$. To make the potential vanish at this minimum, one can add a constant, so that we are left with
\begin{gather}
V(\phi)\propto b\log^2\frac{\phi}{M},
\label{10}
\end{gather} 
with $b$ being a constant. Also, they have considered that the mass $M$ is of order of the GUT (Grand Unified Theories) scale, around the Planck mass. 

\begin{figure*}
	\centering
	\includegraphics[width=7.5cm]{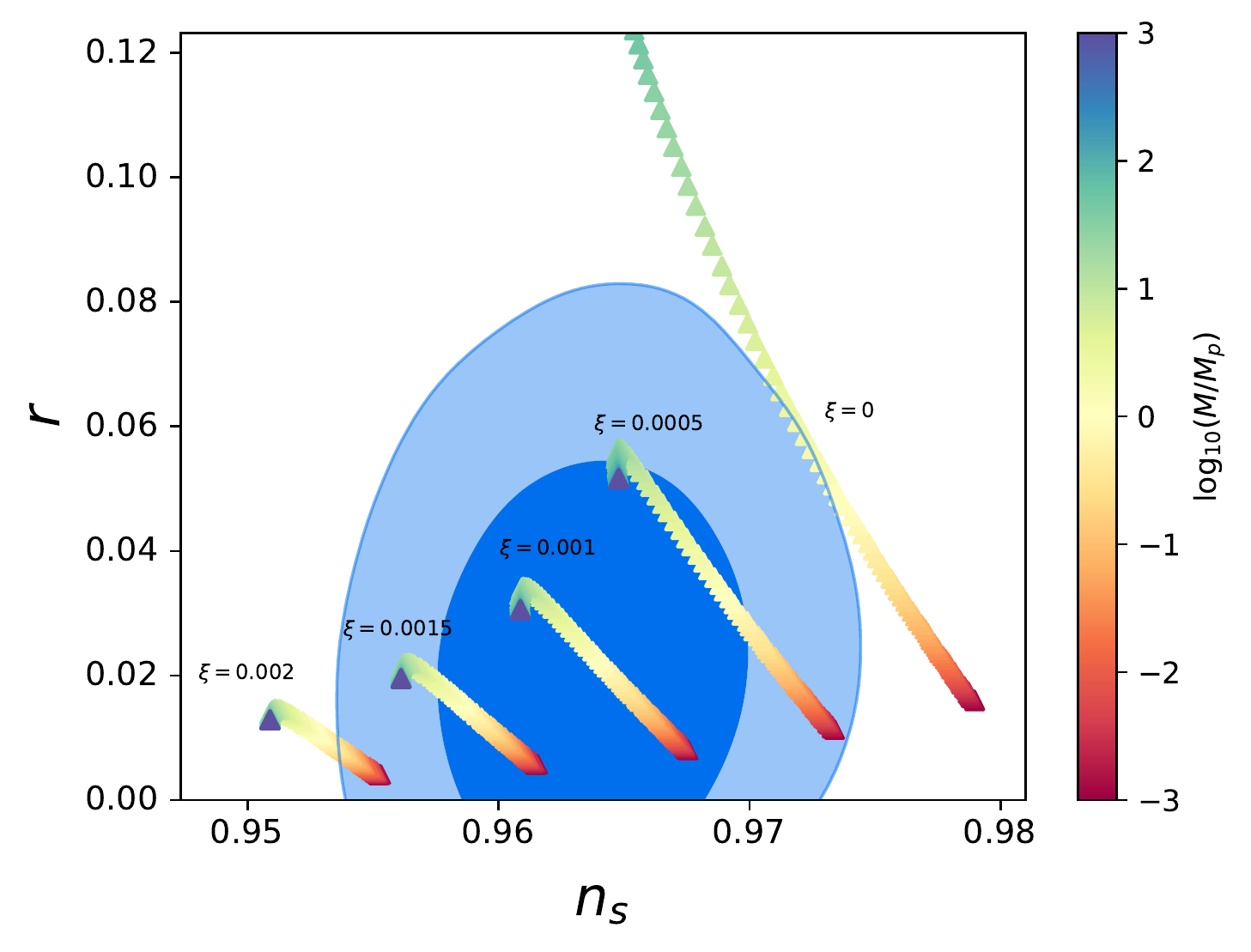}
	\includegraphics[width=7.5cm]{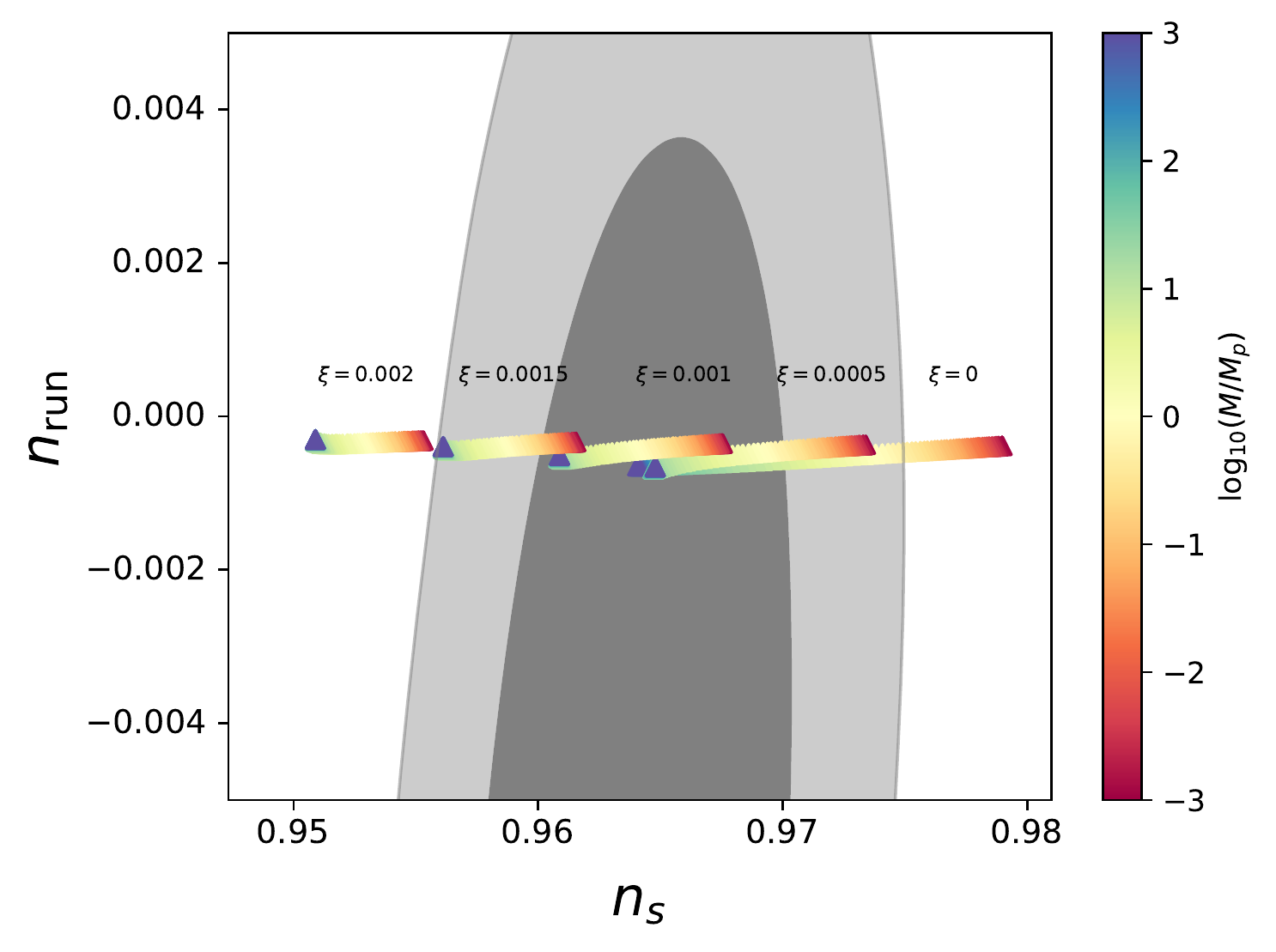}	
	\caption{The $n_s-r$ (left) and $n_s-n_{run}$ (right) predictions for the non-minimally coupled WR model, with the constraints from Planck 2018+BICEP2/Keck. The color bar shows the values of $\operatorname{log}_{10}(M/M_p)=[-3:3]$, while we choose fixed values of $\xi$, indicated in the picture. We also fix $N_\star=55$.}
	\label{fig2}
\end{figure*}

As discussed, while the canonical form $K=T\bar T$ is not very suitable for inflationary purposes, because the resulting potential becomes very steep because of the prefactor $e^K$, the Kähler potential \cite{Kawasaki:2000yn}
\begin{equation}
K_\pm=\pm\frac{1}{2}\left(T\pm \bar T\right)^2 + S\bar S,
\label{11}
\end{equation}
allows the fields to have canonical kinetic terms, while providing a flat direction in which inflation can take place. In \cite{Artymowski:2019jlh}, a more general expression is used by setting $T\rightarrow f(T)$, with the following Kähler function
\begin{equation}
G(T,\bar T) = K(T,\bar T) + \operatorname{log}|W(T)|^2,
\label{12}
\end{equation}
in which the superpotential $W$ is written as
\begin{equation}
W=\Lambda Sw(T),
\label{13}
\end{equation}
so by a choice of $w(T)$, and a redefinition of the field, it is possible to obtain a flat direction where inflation will happen, so that we have the potentials 
\begin{equation}
V_+ = \Lambda^2 w\left(f^{-1}\left(iu_I/\sqrt{2}\right)\right)^2,
\label{14}
\end{equation}
\begin{equation}
V_- = \Lambda^2 w\left(f^{-1}\left(iu_R/\sqrt{2}\right)\right)^2,
\label{15}
\end{equation}
where $f(T)\equiv U=\frac{1}{\sqrt{2}}\left(u_R+iu_I\right)$, and we want to achieve inflation along the real part of the field, $u_R$. We refer to \cite{Artymowski:2019jlh} for the details on the construction of the general potential; for our purposes, we review the functions that lead to the $\log^2(\phi/M)$ function. 
For the $K_-$ and $V_-$ branches, if one chooses
\begin{equation}
f=\frac{M}{\sqrt{2}}e^T, \quad w=T,
\label{16}
\end{equation} 
the following potential is derived
\begin{equation}
V=\Lambda^2e^{u_I^2}\left[\frac{1}{4}\operatorname{log}^2\left(\frac{u_R^2 + u_I^2}{M^2}\right) + \operatorname{arctan}^2\left(\frac{u_I}{u_R}\right)\right],
\label{17}
\end{equation}
which possesses a flat direction along $u_I=0$, and has minima at $u_R=\pm M$; therefore we take $u_R=\phi$, so that the final potential becomes
\begin{equation}
V(\phi)=V_0\operatorname{log}^2\left(\frac{\phi}{M}\right),
\label{18}
\end{equation} 
where $\Lambda^2\equiv V_0$. This method allows us to obtain some well known potentials, such as monomial and Starobinsky ones, while considering a simple form of the superpotential $W$. We note that while in the original scenario that led to (\ref{18}) \cite{Albrecht:1983ib}, the mass $M$ is taken as $M\simeq M_p$, now we have $M$ as a free parameter that can be constrained by observations. In the following discussion, we will investigate the consequences of varying both $M$ and $\xi$, and see how this affects the general predictions.

\subsection{Slow-roll analysis}

We now proceed with the slow-roll prediction of the non-minimally coupled WR inflation. The potential (\ref{18}), in the Einstein frame, becomes

\begin{equation}
\hat{V}(\phi)=\frac{V_0\log^2\left(\frac{\phi}{M}\right)}{\left(1+\xi\frac{\phi^2}{M_p^2}\right)^2},
\label{19}
\end{equation}
shown in Figure \ref{fig1}, for different $\xi$.  From (\ref{19}), using Eq. (\ref{4}) we compute the slow-roll parameters from (\ref{5}) as
\begin{equation}
\epsilon = \frac{2\left[2\left(\frac{\phi}{M_p}\right)^2\log\left(\frac{\phi}{M}\right)\xi-\left(\frac{\phi}{M_p}\right)^2\xi-1\right]^2}{\left(\frac{\phi}{M_p}\right)^2\log^2\left(\frac{\phi}{M}\right)\left(\xi\left(\frac{\phi}{M_p}\right)^2(1+6\xi)+1\right)},
\label{20}
\end{equation}	
\begin{align}
\eta&=\frac{2}{\left(\frac{\phi}{M_p}\right)^2\log^2\left(\frac{\phi}{M}\right)\left(\xi\left(\frac{\phi}{M_p}\right)^2(1+6\xi)+1\right)^2}\Bigg[48\left(\frac{\phi}{M_p}\right)^6\log\left(\frac\phi M\right)^2\xi^4 - 48\left(\frac{\phi}{M_p}\right)^6\log\left(\frac\phi M\right)\xi^4  + 6\left(\frac{\phi}{M_p}\right)^6\xi^4  \nonumber \\ & + 8\left(\frac{\phi}{M_p}\right)^6\log\left(\frac\phi M\right)^2\xi^3 - 8\left(\frac{\phi}{M_p}\right)^6\log\left(\frac\phi M\right)\xi^3 - 60\left(\frac{\phi}{M_p}\right)^4\log\left(\frac\phi M\right)\xi^3 + \left(\frac{\phi}{M_p}\right)^6\xi^3 + 12\left(\frac{\phi}{M_p}\right)^4\xi^3 \nonumber \\ &  + 6\left(\frac{\phi}{M_p}\right)^4\log\left(\frac\phi M\right)^2\xi^2  - 17\left(\frac{\phi}{M_p}\right)^4\log\left(\frac\phi M\right)\xi^2 - 12\left(\frac{\phi}{M_p}\right)^2\log\left(\frac\phi M\right)\xi^2 + 3\left(\frac{\phi}{M_p}\right)^4\xi^2 +  6\left(\frac{\phi}{M_p}\right)^2\xi^2 \nonumber \\ & - 2\left(\frac{\phi}{M_p}\right)^2\log\left(\frac\phi M\right)^2\xi - 10\left(\frac{\phi}{M_p}\right)^2\log\left(\frac\phi M\right)\xi + 3\left(\frac{\phi}{M_p}\right)^2\xi - \log\left(\frac\phi M\right)+1\Bigg].
\label{21}
\end{align}

\begin{figure*}[t]
	\centering
	\includegraphics[width=7.5cm]{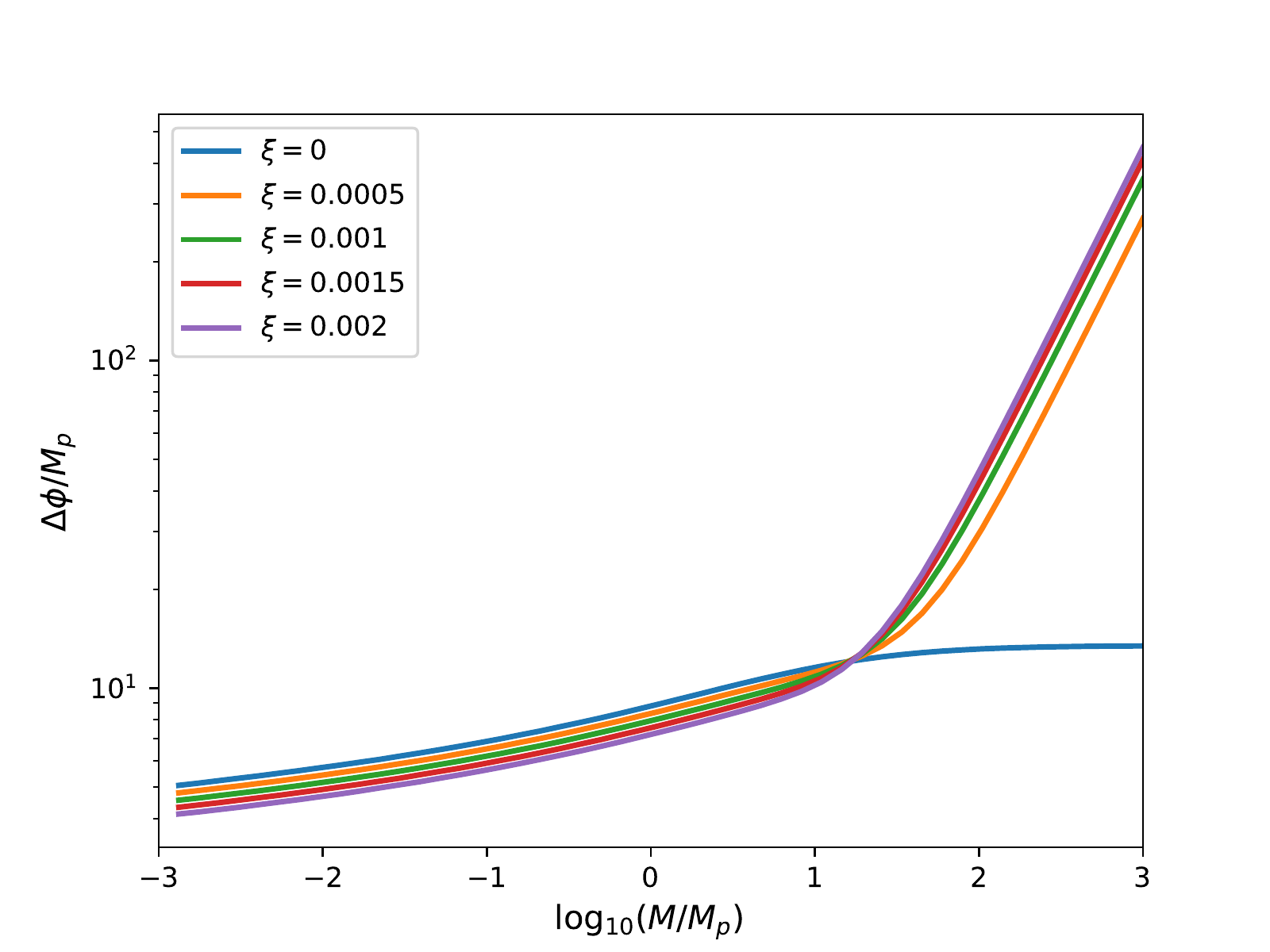}
	\includegraphics[width=7.5cm]{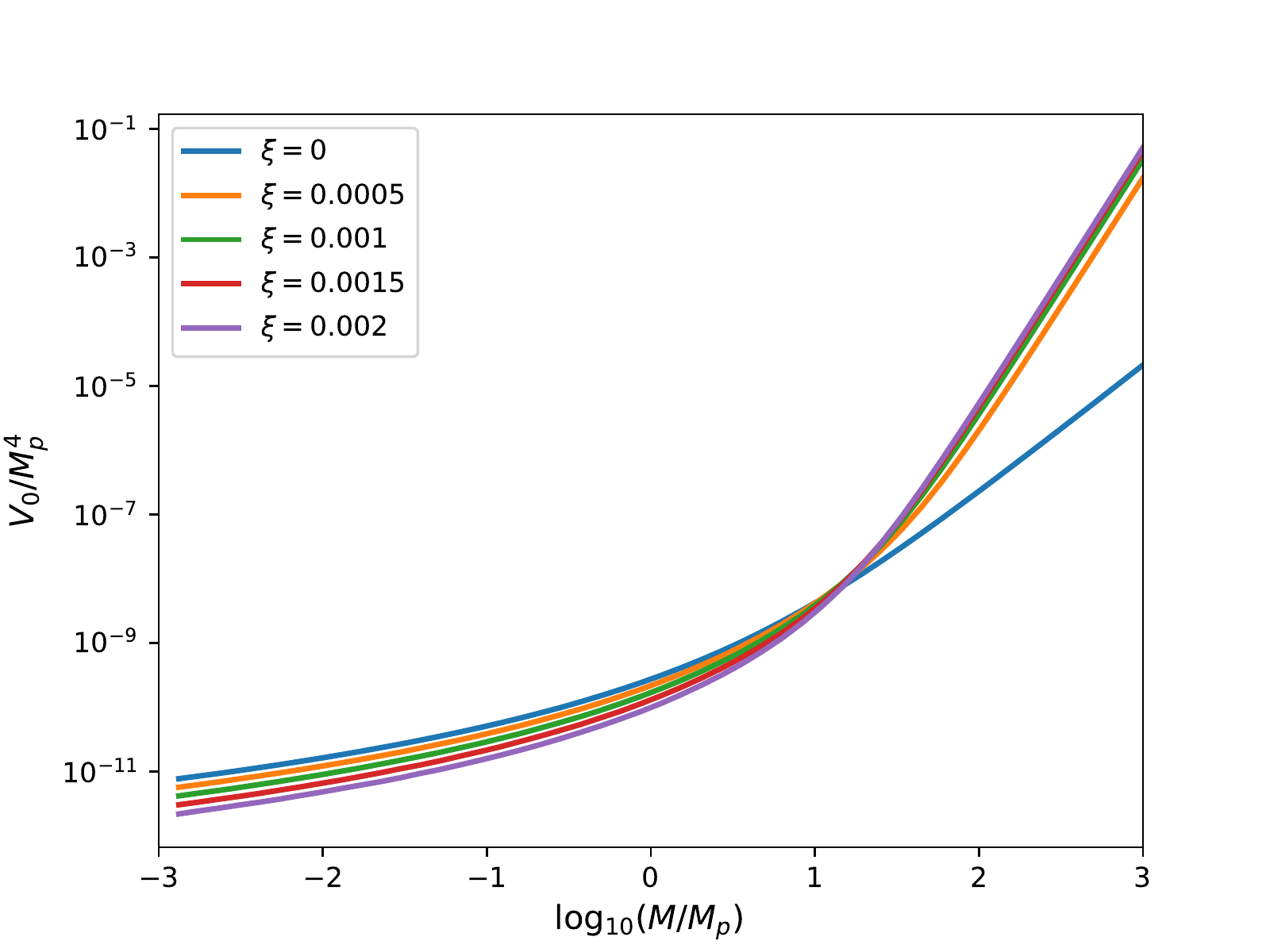}	
	\caption{The field difference $\Delta\phi$ for the WR model as a function of $\operatorname{log}_{10}(M/M_p)$ (left), and the potential amplitude $V_0$ also as a function of $\operatorname{log}_{10}(M/M_p)$ (right), all for selected values of $\xi$.}
	\label{fig3}
\end{figure*}
Inflation might end for either $\epsilon=1$ or $|\eta|=1$, depending on the $M$ and $\xi$ chosen, so we incorporate this feature in our calculations. To compute the field at horizon crossing, we use the expression for the number of e-folds (\ref{6}), which takes the form

\begin{equation}
N_\star = \frac{1}{M_p^2}\int^{\phi_{end}}_{\phi_{\star}}d\phi
\frac{ \frac{\phi}{M_p} \log{\left( \frac{\phi}{M}\right) } \left( \frac{\phi^2}{M_p^2} \xi(1+6\xi)+1\right) }{2 \left( \frac{\phi^2}{M_p^2} \xi+1\right)  \left(  \frac{\phi^2}{M_p^2}\xi \left(2\log{\left( \frac{\phi}{M}\right) }-1\right) \xi-1\right) }
\label{22}
\end{equation}

so we can obtain the predictions for the $n_s$ and $r$ parameters. For a non-zero $\xi$, they read as

\begin{equation}
r = \frac{32\left[2\left(\frac{\phi_\star}{M_p}\right)^2\log\left(\frac{\phi_\star}{M}\right)\xi-\left(\frac{\phi_\star}{M_p}\right)^2\xi-1\right]^2}{\left(\frac{\phi_\star}{M_p}\right)^2\log^2\left(\frac{\phi_\star}{M}\right)\left(\xi\left(\frac{\phi_\star}{M_p}\right)^2(1+6\xi)+1\right)},
	\label{23}
\end{equation}
\begin{align}
n_s-1 & =	\frac{-4}{\left(\frac{\phi_\star}{M_p}\right)^2\log^2\left(\frac{\phi_\star}{M}\right)\left(\xi\left(\frac{\phi_\star}{M_p}\right)^2(1+6\xi)+1\right)^2}\Bigg[24\left(\frac{\phi_\star}{M_p}\right)^6\log^2\left(\frac{\phi_\star}{M}\right)\xi^4 - 24\left(\frac{\phi_\star}{M_p}\right)^6\log\left(\frac{\phi_\star}{M}\right)\xi^4 \nonumber \\ & + 12\left(\frac{\phi_\star}{M_p}\right)^6\xi^4 + 4\left(\frac{\phi_\star}{M_p}\right)^6\log^2\left(\frac\phi M\right)\xi^3 - 4\left(\frac{\phi_\star}{M_p}\right)^6\log\left(\frac{\phi_\star}{M}\right)\xi^3  
- 12\left(\frac{\phi_\star}{M_p}\right)^4\log\left(\frac{\phi_\star}{M}\right)\xi^3  \nonumber \\ & + 2\left(\frac{\phi_\star}{M_p}\right)^6\xi^3 + 24\left(\frac{\phi_\star}{M_p}\right)^4\xi^3 + 6\left(\frac{\phi_\star}{M_p}\right)^4\log^2\left(\frac{\phi_\star}{M}\right)\xi^2  
- 7\left(\frac{\phi_\star}{M_p}\right)^4\log\left(\frac{\phi_\star}{M}\right)\xi^2 \nonumber\\ & + 12\left(\frac{\phi_\star}{M_p}\right)^2\log\left(\frac{\phi_\star}{M}\right)\xi^2 + 6\left(\frac{\phi_\star}{M_p}\right)^4\xi^2 + 12\left(\frac{\phi_\star}{M_p}\right)^2\xi^2
+ 2\left(\frac{\phi_\star}{M_p}\right)^2\log^2\left(\frac{\phi_\star}{M}\right)\xi \nonumber\\ & - 2\left(\frac{\phi_\star}{M_p}\right)^2\log\left(\frac{\phi_\star}{M}\right)\xi + 6\left(\frac{\phi_\star}{M_p}\right)^2\xi + \log\left(\frac{\phi_\star}{M}\right)+2\Bigg].
\label{24}
\end{align}

The $n_s-r$ plane is shown in Figure \ref{fig2} (left). We choose values of $\xi$ in the interval $\xi\in[0,0.002]$, while considering a wide mass range $\log_{10}(M/M_p)\in[-3,3]$, for $N_\star=55$. On the right side of the figure, we see the curve marginally in the $2\sigma$ region corresponding to the minimally coupled model; as $M$ increases, $r$ grows significantly, up to a limit that depends on $N_\star$. For smaller $M$, however, $r$ becomes consistent with data, but $n_s$ does not, taking larger values. We then realize that for $N_\star=50-60$, the minimally coupled model does not agree with Planck data at the $1\sigma$ level, regardless of which $M$ we take. When we consider a non-minimal coupling, however, the picture changes completely. Even for a small coupling, the predictions of both $n_s$ and $r$ enter the $1\sigma-2\sigma$ region easily for all ranges of $M$ considered; this means that the upper limit on $r$ is different for each $\xi$ chosen; in fact, we see that at some point $r$ starts decreasing as $M$ grows further. The upper limit on $\xi$ we show in the plot is $\xi=0.002$, since the predictions start to leave the contours for larger values; we see then that the model is consistent with a small non-minimal coupling of order $\xi\sim 10^{-3}$.

Also in Figure \ref{fig2} (right), we show the predictions for the running of the spectral index $n_{run}$. Planck restricts its value as $n_{run}=-0.0045\pm0.0067$ at $68\%$ confidence level \cite{Planck:2018jri}, and the Planck+TT +TE+EE+lowE+lensing constraints are also shown. The WR model predicts values that are consistent with data, where the parameter $M$ seems to have a more significant impact on the values of $n_{run}$, since, as $M$ decreases, the running increases towards positive values.

Figure \ref{fig3} (left) shows the excursion of the field $\Delta\phi\equiv\phi_{\star}-\phi_{end}$ as a function of the mass $M$. For the minimally coupled model, $\Delta\phi$ increases with $M$ but tends to a roughly constant value for very large $M$; the non-minimally coupled model mostly follows the same behavior for smaller $M$, but we see that for around $M\geq 10$, $\Delta\phi$ increases significantly, so that inflation happens for a large interval of $\phi$. Note that this corresponds to the \textquoteleft turning point'  for $r$ as, from the same limit, $r$ starts decreasing in the $n_s-r$ plot of Figure \ref{fig2}. We also see that for the interval of $M$ considered, the field excursion is always larger than one, but as $\xi$ increases, $\Delta\phi/M_p$ decreases slightly. On the right figure (in Figure \ref{fig3}), we see that the amplitude of the potential $V_0$, which is calculated from

\begin{align}
\frac{V_0}{M_p^4} & = \frac{24\pi^2 P_R \left(1+\xi\left(\frac{\phi_\star}{M_p}\right)^2\right)^2}{\log^2\left(\frac{\phi_\star}{M}\right)\left(\frac{\phi_\star}{M_p}\right)^2\left(\xi\left(\frac{\phi_\star}{M_p}\right)^2(1+6\xi) + 1\right)}\Bigg[\Bigg(4\log^2\left(\frac{\phi_\star}{M}\right) -4\operatorname{log}\left(\frac{\phi_\star}{M}\right)+ 1\Bigg)2\xi^2\left(\frac{\phi_\star}{M_p}\right)^4 \nonumber\\ &  + \left(1-2\operatorname{log}\left(\frac{\phi_\star}{M}\right)\right)4\xi\left(\frac{\phi_\star}{M_p}\right)^2 + 2\Bigg]
\label{25}
\end{align}

increases by many orders of magnitude with $M$, especially in the presence of the non-minimal coupling, having similar values for all $\xi\neq0$ considered.

\section{Pertubative reheating and connection with CMB parameters}\label{sec4}

\begin{figure*}[t]
	\centering
	\includegraphics[width=7.5cm]{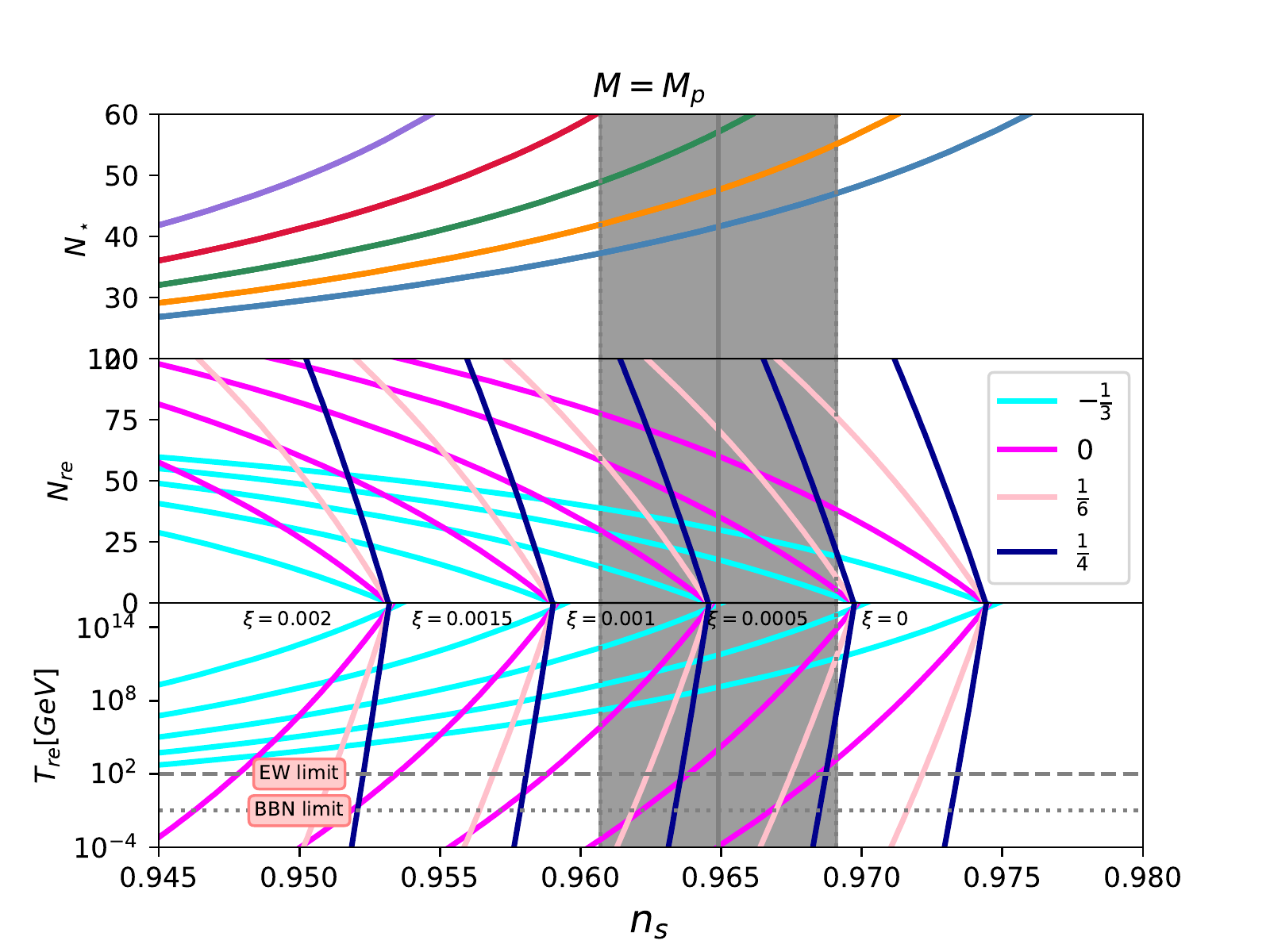}
	\includegraphics[width=7.5cm]{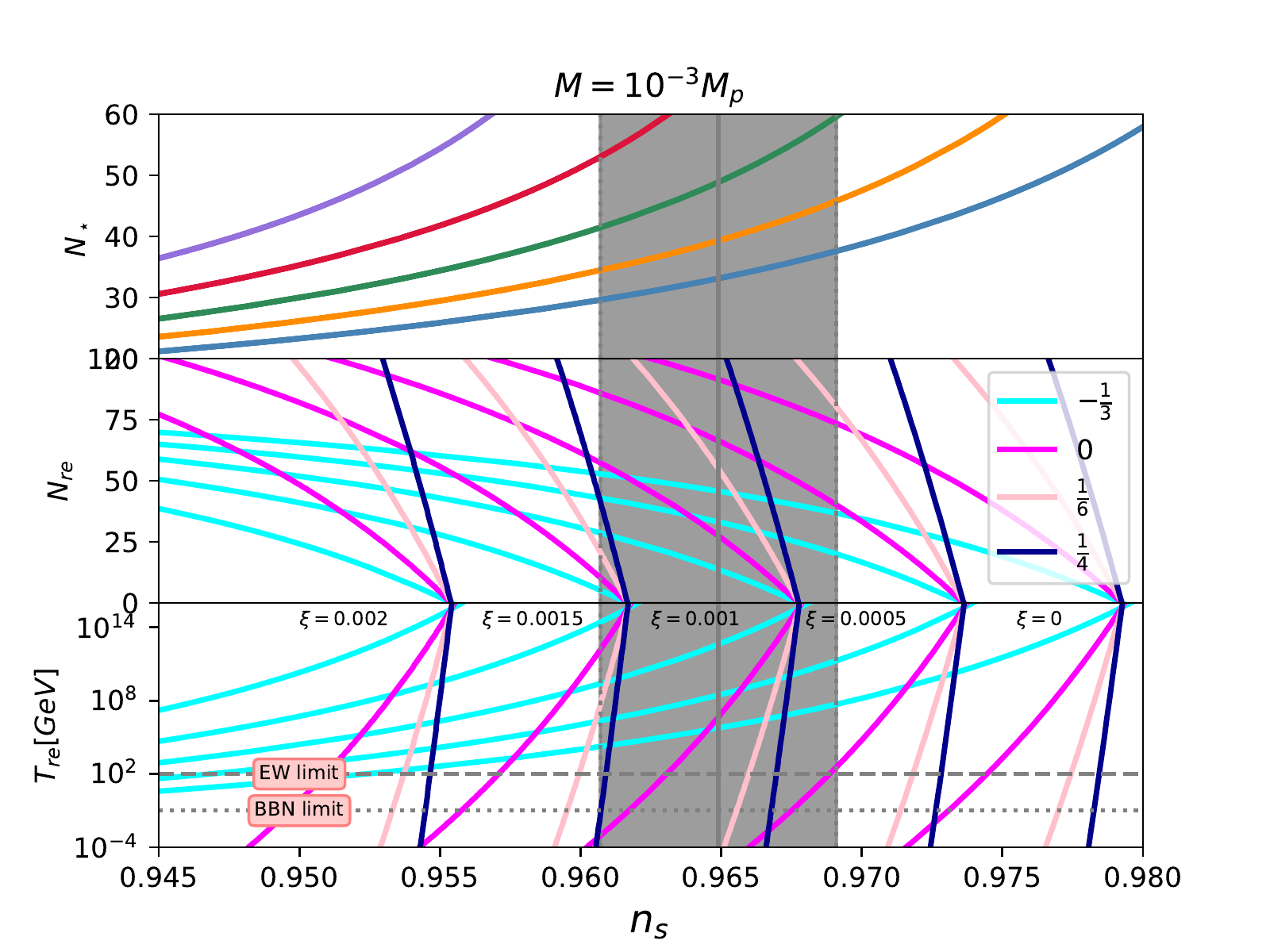}
	\caption{The number of e-folds $N_k$, and the predictions for $N_{re}$ and $T_{re}$ given by the model in Eq. (\ref{19}). We have fixed $M=M_p$ (left) and $M=10^{-3}M_p$ (right), and have chosen values of $\xi$ that cover all interval considered, of $\xi=0,0.005,0.001,0.0015,0.002$, whose colors correspond to the values indicated in Figures \ref{fig1} and \ref{fig3}. Also, four values of $\bar w_{re}$ are chosen: $\bar w_{re}=-1/3$ (cyan line), $\bar w_{re}=0$ (magenta line), $\bar w_{re}=1/6$ (pink line) and $\bar w_{re}=1/4$ (dark blue line). The grey vertical band represents the $1\sigma$ constraint on $n_s$ from Planck TT,TE,EE+lowE+lensing, while the horizontal dashed/dotted grey lines represent the electroweak and BBN bounds respectively.}
	\label{fig4}
\end{figure*}

One of the predictions of slow-roll inflation caused by one (or more) scalar field(s), is that after the accelerated expansion ends, the \textit{inflaton} decays into the Standard Model particles to realize the Hot Big Bang scenario, starting the radiation-dominated period of cosmic evolution. Such process is called \textit{reheating}, and a variety of mechanisms for this process exist, such as perturbative reheating \cite{Abbott:1982hn,Albrecht:1982mp}, parametric resonance \cite{Greene:1997fu,Kofman:1994rk,Kofman:1997yn} and tachyonic resonance \cite{Greene:1997ge,Dufaux:2006ee,Abolhasani:2009nb}. Still, the physics of reheating is not well known, and one must seek indirect ways of probing this epoch and put theoretical and observational constraints. An approach for investigating the reheating period by connection with a given model was used in several works \cite{Munoz:2014eqa,Cook:2015vqa,Cai:2015soa,Dai:2014jja,Ueno:2016dim,Eshaghi:2016kne,Kabir:2016kdh,DiMarco:2017zek,Drewes:2017fmn,Lopez:2021agu,Cheong:2021kyc}, in a way that makes it possible to link post-inflationary quantities with CMB parameters, such as $n_s$, $r$ and $N_k$. We follow the approach described in these works, which we review here. 

Assuming that the energy density during reheating evolves as $\rho\propto a^{-3(1+w_{re})}$, we write the ratio $\rho_{end}/\rho_{re}$ as
\begin{gather}
\frac{\rho_{end}}{\rho_{re}}=\left(\frac{a_{end}}{a_{re}}\right)^{-3(1+w_{re})}
\label{26}
\end{gather}
with $\rho_{end}$ being the energy density at the end of inflation, and $w_{re}$ is the equation of state parameter; it is possible to consider a parameterization for $w_{re}$ to account for the transition between inflation and radiation era \cite{Saha:2020bis,DiMarco:2021xzk}, however, following previous works, here we use the average $\bar{w}_{re}$ as a constant. Now we assume two features: First, that reheating ends when $H=\Gamma$, with $\Gamma$ being a dissipation coefficient determined by the inflaton coupling with other fields, and second, that the energy density can be expressed as
\begin{equation}
\rho_{re} = \frac{\pi^2}{30}g_{re}T_{re}^4,
\label{27}
\end{equation}
at reheating, where $g_{re}$ is the number of relativistic degrees of freedom, generally assumed as being $g_{re}\sim 10^2$. Using the definition of the number of e-folds $N$, we write
\begin{equation}
N_{re} =\ln\left(\frac{a_{re}}{a_{end}}\right) = \frac{1}{3(1+\bar{w}_{re})}\ln\left(\frac{\rho_{end}}{\rho_{re}}\right).
\label{28}
\end{equation} 
Since $\rho_{end}=\frac{3}{2}V(\phi_{end})$, from the definition of the equation of state parameter, obtained by setting $w_{end}=-1/3$ at the end of inflation, we have
\begin{equation}
N_{re}=\frac{1}{3(1+\bar{w}_{re})}\ln\left(\frac{45V_{end}}{\pi^2g_{re}T^4_{re}}\right).
\label{29}
\end{equation}
Assuming that entropy is conserved from the end of reheating until today \footnote{This means that $gT^3a=const.$ so we can relate the reheating era with the present time as \begin{equation}
	g_{re}T_{re}^3 = \Big(\frac{a_{0}}{a_{re}}\Big)^3\Big( 2T_{0} + \frac{21}{4}T_{\nu,0}^3\Big),
	\label{30}
	\end{equation} where $T_{\nu,0}$ is the neutrino temperature, related with the photon one as $T_{\nu,0}=(4/11)T_0$.}, one can obtain a relation between the present CMB temperature and the temperature of the thermal bath at the end of reheating (also called \textquoteleft reheating temperature')
\begin{equation}
\frac{T_{re}}{T_0}=\frac{a_{0}}{a_{eq}}e^{N_r}\left(\frac{43}{11g_{re}}\right)^{1/3},
\label{31}
\end{equation}
with $N_r$ being the number of e-folds of the radiation era, $T_0$ is the current CMB temperature and $a_{eq}$ is the scale factor at matter-radiation equality. As the ratio $a_0/a_{eq}$ can be written as $a_0/a_{eq}=a_0H_ke^{-N_k}e^{-N_{re}}e^{-N_r}/k$, we can rewrite (\ref{31}) as
\begin{equation}
T_{re}=\left(\frac{43}{11g_{re}}\right)^{1/3}\left(\frac{a_0T_0}{k}\right)H_ke^{-N_k}e^{-N_{re}},
\label{32}
\end{equation}
obtaining an important relation between $T_{re}$ and $N_{re}$. Large temperatures correspond to a more efficient reheating, causing the duration to decrease. This way, if $N_{re}\rightarrow0$, we find a limit for $T_{re}$ that corresponds to the \textit{maximum reheating temperature}, achieved for a instantaneous reheating. 
Back to (\ref{29}), if we substitute (\ref{32}) into it, we will eventually get to
\begin{equation}
N_{re}   = \frac{4}{3\bar{w}_{re}-1}\Bigg[N_k + \ln\left(\frac{k}{a_0T_0}\right) + \frac{1}{4}\ln\left(\frac{45}{g_{re}\pi^2}\right) + \frac{1}{3}\ln\left(\frac{11g_{re}}{43}\right) + \ln\left(\frac{V_{end}^{1/4}}{H_k}\right)\Bigg]
\label{33}
\end{equation}

\begin{figure*}[t]
	\centering
	\includegraphics[width=7.5cm]{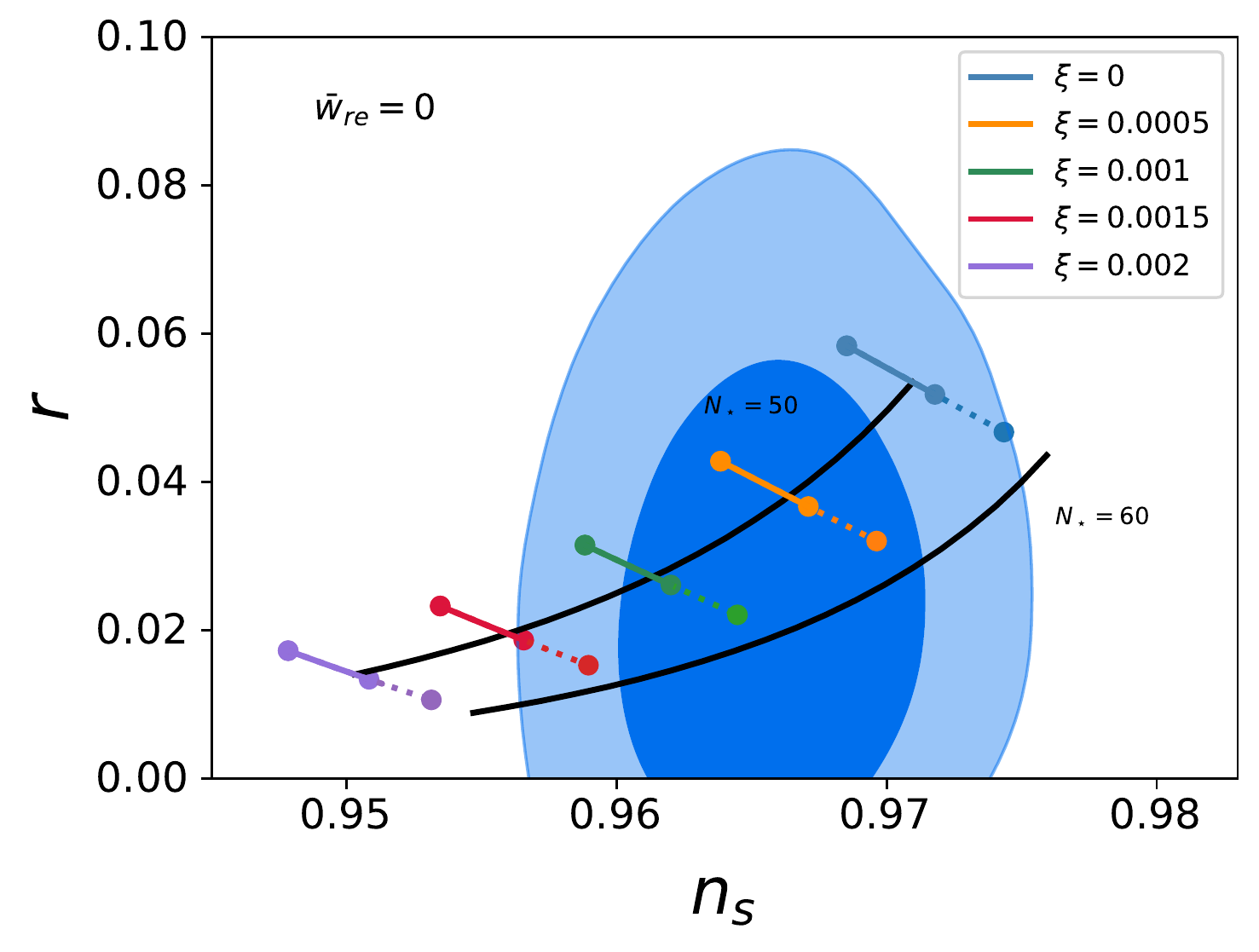}
	\includegraphics[width=7.5cm]{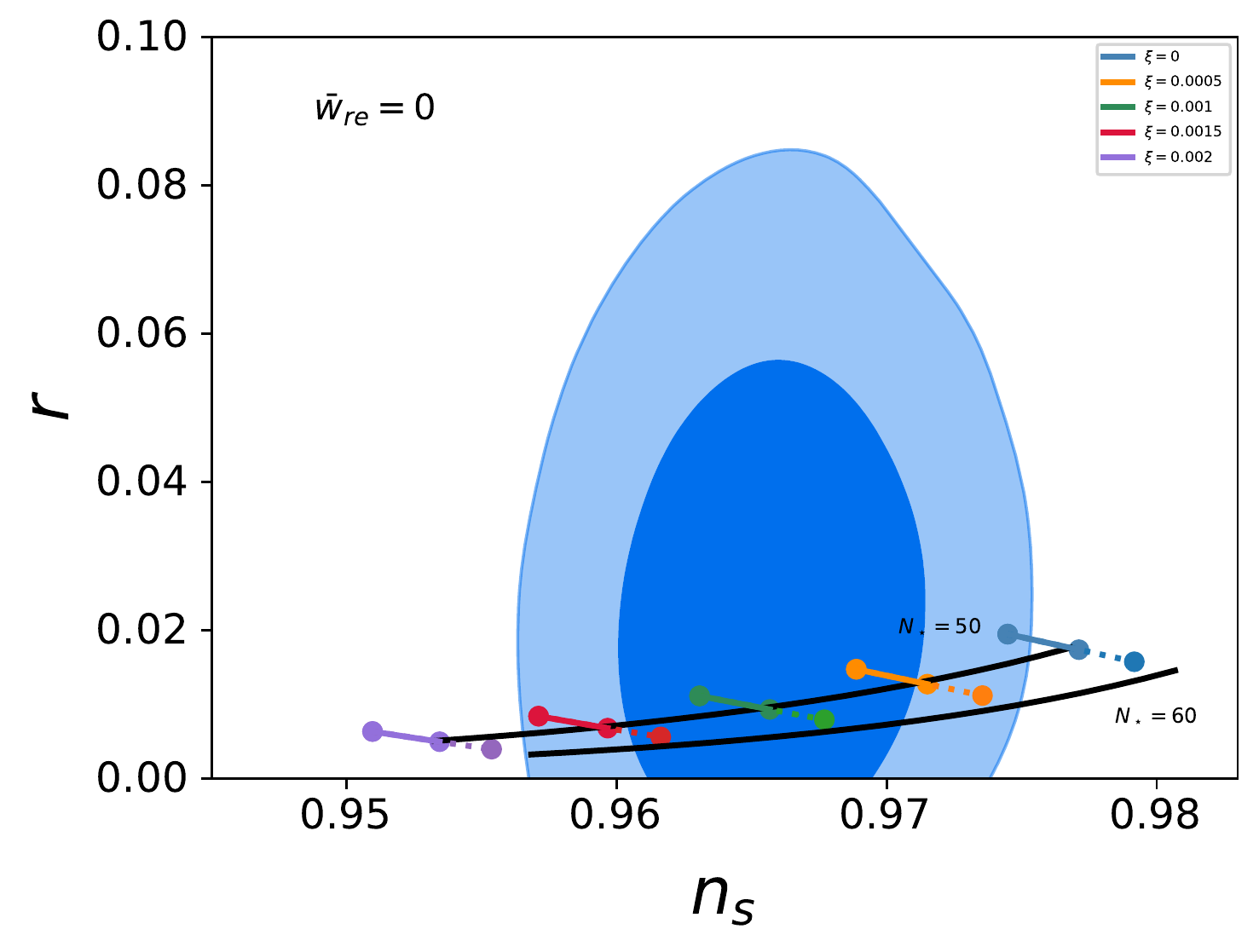}	
	\includegraphics[width=7.5cm]{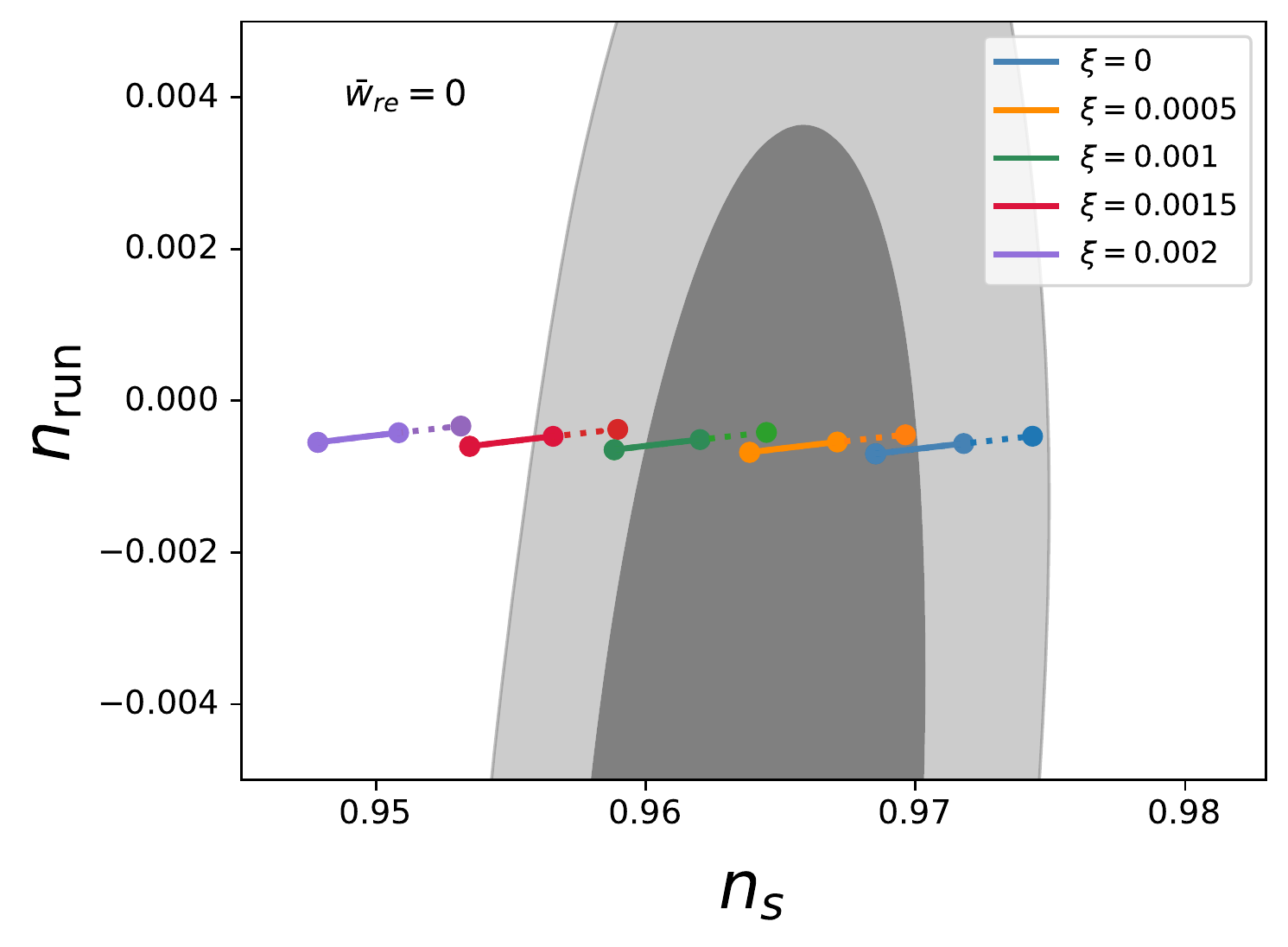}	
	\includegraphics[width=7.5cm]{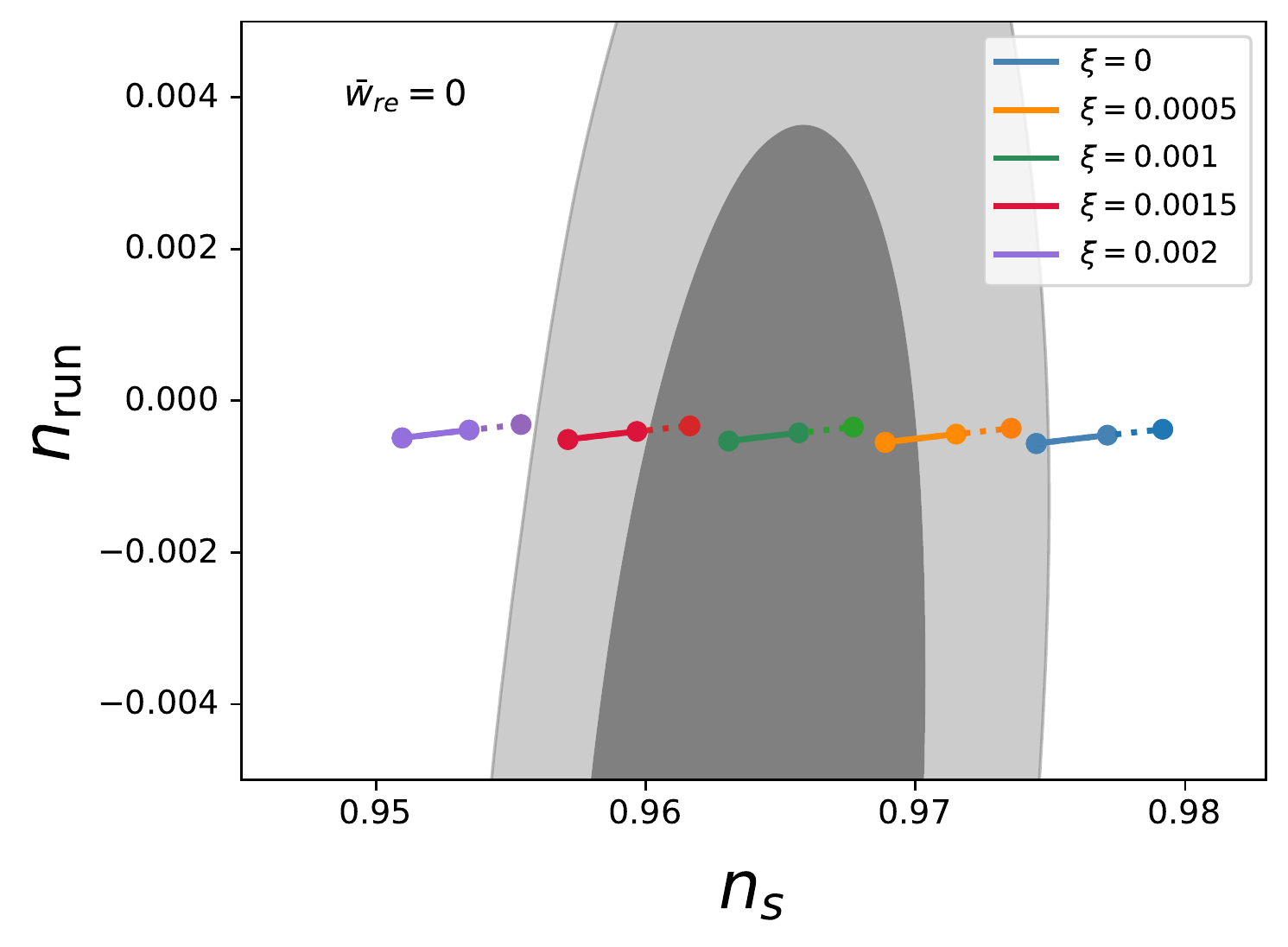}				
	\caption{The $n_s-r$ and $n_s-n_{run}$ planes for the model (\ref{19}), where now we take into account the limits on the reheating temperature $T_{EW}\leq T_{re}\leq 10^9GeV$, for $\bar w_{re}=0, M=M_p$ (left) and $M=10^{-3}M_p$ (right), with each color corresponding to a different $\xi$, as indicated. The segments with dotted lines correspond to the limit $10^9GeV<T_{re}<T_{max}$. For reference, we also show the curves for a varying $\xi$, when $N_\star=50,60$, for the $n_s-r$ values (black lines).}
	\label{fig5}
\end{figure*}

Eqs. (\ref{32}) and (\ref{33}) are the resulting expressions of this approach. The main thing we notice here is that one only needs to choose a specific model as input, so that we can derive constraints on the reheating era as a function of the model parameters \footnote{We also choose $k/a_0=0.05$Mpc$^{-1}$, $T_0=2.725$K and $g_{re}=200$.}. We also note that a series of assumptions were made here for these expressions to be achieved; we are not considering the specifics of couplings of the inflaton to another fields, so other processes, like parametric resonance right at the first oscillations of the field around the minimum of the potential are neglected, and have to be investigated in a different manner. Yet we still can have an estimate for the temperature and duration of reheating in this context, as most of the reheating period is usually considered as characterized by the perturbative decay of the inflaton.

This is shown in Figure \ref{fig4}, for the non-minimally coupled WR model. We have plotted the number of e-folds of inflation $N_k$, the reheating temperature $T_{re}$ and its duration $N_{re}$ both as a function of the spectral index $n_s$, where it is also possible to see that we have considered different values of $\xi$, while keeping $M$ fixed. Before proceeding, we should remember that there are both lower and upper well motivated limits on $T_{re}$, given by the Big Bang Nucleosynthesis (BBN) and the case of instantaneous reheating, where $N_{re}=0$. As the BBN limit can be taken as $T_{BBN}\sim 10$ MeV, we must set $T_{re}$ to at least a value that is higher than that, such that there is no interference in the future creation of atoms. This limit is expressed by the dotted horizontal grey line in Figure \ref{fig4}. Also, for reference, we show the electroweak scale $T_{EW}=100$ GeV, expressed by the dashed horizontal grey line, and it will be the lower limit for $T_{re}$ adopted here. The upper limit on $T_{re}$ comes from setting $N_{re}=0$, and it is also represented directly by the choice $w_{re}=\frac{1}{3}$, meaning that the universe goes instantly from an inflationary to a radiation-dominated regime. Although not shown, this would appear in Figure \ref{fig4} as a vertical line that passes through the convergence point of all other lines; this means that the instantaneous reheating limit gives a unique prediction on $n_s$. 

For $M=M_p$ (left figure), we see that as $\xi$ increases, the curves all shift to the left, in accordance with the $n_s-r$ plane shown in Figure \ref{fig2}. In particular, for $\xi=0$ (the minimally-coupled case) a quite long period of reheating is allowed, up to $N_{re}\sim50$, if we take $T_{BBN}$ as the lower limit. For different $\bar w_{re}$, one can see how the curves change so there can be a wider range of $N_{re}/T_{re}$ allowed by data at $1\sigma$ level, if we take the $\xi=0.0005$ case as an example. We note a similar behavior for $M=10^{-3}M_p$ (right figure), with the difference that lower $M$ leads to a higher $n_s$, so that the $\xi=0$ case has little concordance with the 1$\sigma$ region of $n_s$ data; looking at the curve for $N_k$ (in blue), this concordance would only happen for a quite low $N_k$, of around $\sim 40$. This means that $\xi\neq 0$ is required in one wants a value for $n_s$ closer to the central one.

A more stringent constraint on $T_{re}$ can be established by the so-called gravitino overproduction problem \cite{Khlopov:1984pf,Kawasaki:1994af,Bolz:2000fu,Kawasaki:2008qe}. This bound comes from supersymmetric derived scenarios in which the excessive production of such particles would be enough to overclose the Universe, depending on their masses. In particular, the decay of high mass gravitinos could interfere in well-established processes, such as the BBN; the highest bound on the reheating temperature that can be set comes from $m_{3/2}\geq 1$TeV, where $m_{3/2}$ is the gravitino mass \cite{Kawasaki:1994af} . This implies $T_{re}\leq10^{9}$GeV, which is the upper bound we will use to create the plots in Figure \ref{fig5}. We have again the $n_s-r$ and $n_s-n_{run}$ plane, but with values that correspond to the limits on $T_{re}$ just discussed. We have fixed $M=M_p$ (left panels) and $M=10^{-3}M_p$ (right panels) with $\bar w_{re}=0$ and considered different values of $\xi$. When taking these constraints into consideration, the minimally-coupled model now can be well within the $2\sigma$ region of the confidence contours, while for $\xi\sim0.0005-0.001$, there is full concordance with the $1\sigma$ region. This is because, as $T_{re}$ grows, so does $N_k$, and consequently the limit $T_{EW}\leq T_{re}\leq 10^9GeV$ corresponds to approximately $N_k\sim 46.2-51.3$ ($M=M_p$) and $N_k\sim 45.4-50.5$ ($M=10^{-3}M_p$), for all values of $\xi$ considered. As for the duration of reheating, for the same parameters, and imposing the limits on the temperature, we obtain $N_{re}\sim 20-40$ for both $M=M_p$ and $M=10^{-3}M_p$ approximately.

\section{Discussion and Conclusions}\label{sec6}

The consideration of the presence of a non-minimal coupling of the inflaton with gravity has been a subject of discussion for a long time, but with the increasingly accurate cosmological measurements that are able to impose severe constraints on many different models, it was realized how important of a role such a coupling can have in making models more consistent with data. In particular, classes of models such as chaotic inflation, initially excluded by data for having a high tensor-to-scalar ratio, now can have its predictions accomodated into the most recent CMB constraints, therefore, making them viable models again. Another example of such model is the Witten-O'Raifeartaigh inflation, derived from a supersymmetric context. In the original discussion, the potential is characterized by a quadratic logarithm form, with a minimum determined by a mass scale $M$ \cite{Albrecht:1983ib}. This parameter is theoretically estimated as roughly the Planck mass, so $M\simeq M_p$; on the other hand, a recently supersymmetric realization of the same potential does not necessarily impose any theoretical constraints on $M$ \cite{Artymowski:2019jlh}. This allows us to investigate the viability of the model by exploring the effect of both $\xi$ and $M$, which is what we have done in this work. For a nonzero $\xi$, we have found that the slow-roll analysis allows a very wide range of the mass scale, as an increasing $M$ no longer results in a tensor-to-scalar ratio outside the Planck upper bound; also, a decreasing $M$, while it still leads to an increasing $n_s$, allows us now to achieve a low $r$ while still being in the $1\sigma-2\sigma$ region of the $n_s-r$ plane. We also see the impact on the running of the scalar index; a non-zero $\xi$ leads the predictions into the $1\sigma$ region of the $n_s-n_{run}$ plane, where the parameter $M$ has the interesting feature of slightly increase or decrease $n_{run}$.

With these results, we have proceeded to investigate the reheating era. It is possible, under certain assumptions, to relate CMB quantities with post-inflationary ones, such as the reheating temperature $T_{re}$ and its duration, characterized by the number of e-folds $N_{re}$, when perturbative reheating takes place \cite{Munoz:2014eqa,Cook:2015vqa,Cai:2015soa,Dai:2014jja,Ueno:2016dim,Eshaghi:2016kne,Kabir:2016kdh,DiMarco:2017zek,Drewes:2017fmn,Lopez:2021agu,Cheong:2021kyc}. Assuming a constant equation of state parameter $\bar w_{re}$ (or a given parameterization) during reheating, one can compute the dependence of $(T_{re},N_{re})$ on $n_s$ or $r$, for instance, as shown in Figure \ref{fig4}. This method has been used for minimally coupled models with monomial potentials and for the $\alpha$-attractor scenario, but recently, applications to non-minimally coupled models have appeared \cite{Cheong:2021kyc,Kawai:2021hvs}, showing how the presence of the coupling might affect the post-inflationary parameters. We have obtained an interval of approximately $20<N_{re}<50$ for the duration of the reheating, when $\bar w_{re}=0$, which also imposes a constraint on $N_\star$ at horizon crossing, as seen in Figure \ref{fig5}. 

The natural path of investigation for this scenario then, is the restriction of cosmological parameters by using the full Planck data, as we must see how the change in the model parameters will affect the temperature power spectrum. This is the subject of a next paper that is in preparation, where we will check for a restriction on the mass scale $M$ and see if there is a preference for a non-minimal coupling, represented by $\xi$.\\

\section*{Acknowledgements}

F.B.M. dos Santos is supported by Coordena\c{c}\~{a}o de Aperfei\c{c}oamento de Pessoal de N\'ivel Superior (CAPES). R. Silva acknowledges financial support from CNPq (Grant No. 307620/2019-0).

\bibliography{references}

\end{document}